\documentclass{aa}
\usepackage{graphicx}
\usepackage{txfonts}
\usepackage{xspace}
\usepackage{xcolor}
\usepackage{url}
\usepackage{hyperref}
\usepackage{orcidlink}
\newcommand {\ixpe}{{IXPE}\xspace}

\newcommand{\vel}{\mbox{Vela~X-1}\xspace}

\begin{document} 

\title{Revealing two orthogonally polarized spectral components \\ in Vela X-1 with IXPE}

\author{%\small 
Sofia~V.~Forsblom \inst{\ref{in:UTU}}\orcidlink{0000-0001-9167-2790}
\and Sergey~S.~Tsygankov \inst{\ref{in:UTU},\ref{in:IKI}}\orcidlink{0000-0002-9679-0793}
\and Valery~F.~Suleimanov \inst{\ref{in:Tub}}\orcidlink{0000-0003-3733-7267}
\and Alexander~A.~Mushtukov \inst{\ref{in:Oxford}}\orcidlink{0000-0003-2306-419X}
\and Juri~Poutanen \inst{\ref{in:UTU},\ref{in:IKI}}\orcidlink{0000-0002-0983-0049}
          }
          
\institute{Department of Physics and Astronomy, FI-20014 University of Turku,  Finland \label{in:UTU} \\ \email{sofia.v.forsblom@utu.fi}
\and 
Space Research Institute, Russian Academy of Sciences, Profsoyuznaya 84/32, Moscow 117997, Russia \label{in:IKI} 
\and
Institut f\"ur Astronomie und Astrophysik, Universit\"at T\"ubingen, Sand 1, D-72076 T\"ubingen, Germany \label{in:Tub}
\and 
Astrophysics, Department of Physics, University of Oxford, Denys Wilkinson Building, Keble Road, Oxford OX1 3RH, UK \label{in:Oxford}
}
          
\titlerunning{Two polarized emission components in \vel}
\authorrunning{S.~V.~Forsblom et al.}

\date{2024}

\abstract{
Polarimetric observations of X-ray pulsars (XRPs) have provided us with the key to unlocking their geometrical properties.
Thanks to the Imaging X-ray Polarimetry Explorer (\ixpe), the geometries of several XRPs have been determined, providing new insights into their emission mechanisms and magnetic field structures.
Previously, \vel has proven to be exceptional in demonstrating a clear energy dependence of its polarimetric properties, showing a 90\degr\ swing in the polarization angle (PA) between low and high energies. 
Due to the complex energy-dependent nature of the polarization properties, it was not possible to determine the pulsar geometry.
In this work, we present the results of a detailed analysis of the pulse phase-resolved polarization properties of \vel at different energies.
By separating the polarimetric analysis into low and high energy ranges, we are able to disentangle the contributions of the soft and hard spectral components to the polarization, revealing the pulse phase dependence of polarization degree (PD) and PA in each energy band.
The PA pulse phase dependence at high energies (5--8 keV) allows us, for the first time, to determine the pulsar geometry in \vel.
The fit with the rotating vector model gives an estimate for the pulsar spin position angle at around 127\degr\ and for the magnetic obliquity of 13\degr.
In order to explain the 90\degr\ swing in PA between high and low energies, we discuss two possible scenarios: a two-component spectral model and the vacuum resonance.
}


\maketitle

\section{Introduction} 
\label{sec:intro}

Accreting X-ray pulsars (XRPs)  are highly-magnetized neutron stars (NSs) accreting matter from massive stellar companions (see \citealt{MushtukovTsygankov2024} for a recent review).
They provide us with indispensable laboratories for studying the processes related to the complex interplay between accreting matter, radiation, and the strong magnetic field of the NS itself.
As a consequence of the strong magnetic field, the accreting matter is confined to small regions around the magnetic poles of the NS, which leads to the appearance of pulsed X-ray emission as the NS rotates.
The strong magnetic field is also the main cause of the highly polarized X-ray emission expected from these sources.
Due to the birefringence of highly-magnetized plasma, the emission can be treated in terms of two normal polarization modes: the ordinary mode (O-mode) and the extraordinary mode (X-mode).
The significant difference in opacity of these two modes (below the cyclotron energy) means that, in theory, a large degree of polarization is expected to be observed from these sources \citep{Meszaros88,2021MNRAS.501..109C}.
Recently, the launch of the Imaging X-ray Polarimetry Explorer (\ixpe) in December 2021 has added a tool to examine the X-ray polarimetric properties of XRPs.
%: X-ray polarimetry.
Importantly for XRPs, their geometrical properties can be determined as a result of the additional information encrypted in the behavior of the polarization properties with the pulsar phase.

Several XRPs have been observed by \ixpe (see recent review by \citealt{Poutanen2024b}): \mbox{Her~X-1} \citep{2022NatAs...6.1433D,Heyl24,2024MNRAS.tmp.1187Z},  \mbox{Cen~X-3} \citep{2022-cenx3}, X~Persei \citep{2023MNRAS.524.2004M}, \mbox{4U~1626$-$67} \citep{2022ApJ...940...70M}, \mbox{Vela~X-1} (\citealt{Forsblom2023}, hereafter Paper I), \mbox{GRO~J1008$-$57} \citep{2023A&A...675A..48T}, EXO~2030+375 \citep{2023A&A...675A..29M}, \mbox{LS~V~+44~17} \citep{2023A&A...677A..57D}, \mbox{GX~301$-$2} \citep{Suleimanov2023}, \mbox{Swift~J0243.6+6124} \citep{Poutanen24}, and \mbox{SMC~X-1} \citep{Forsblom24}. 
Their polarization properties show a pattern of rather low ($\sim$10\%) linear polarization degree (PD), even in the pulse phase-resolved polarimetric data, putting earlier theoretical predictions into question.
Additionally, the geometrical properties of the majority of XRPs observed by \ixpe have been determined by modeling the pulse phase dependence of the polarization angle (PA) with the rotating vector model \citep[RVM;][]{Radhakrishnan69,Meszaros88,Poutanen2020}.

\ixpe observations of the wind-accreting high-mass X-ray binary (HMXB) \vel (also known as 4U 0900$-$40) did indeed reveal a relatively low phase-averaged PD of $2.3\pm0.4\%$ with a PA of $-47\fdg3\pm5\fdg4$ in the 2--8 keV energy band (Paper I).
The phase-averaged energy-resolved polarimetric analysis provided the most interesting results, revealing a clear energy dependence of the polarization properties in the \ixpe 2--8 keV energy band, with a swing of the PA by $90\degr$ at around 3.4 keV, a unique behavior among the XRPs observed by \ixpe. 
The PD shows a decrease from around 4\% at 2--3 keV, down to zero at 3.4~keV, then increasing to about 10\% at 7--8 keV with a PA rotated by $\sim90\degr$ relative the low-energies.
The phase-resolved polarimetric analysis also found a PD in the range 0--9\% with the PA varying between $-80\degr$ and~$40\degr$.
However, the complicated nature of the PA pulse-phase dependence cannot be explained using the RVM, and, therefore, the pulsar geometry could not be determined for \vel.

It is important to note that additional polarized contributions may arise in the system, for example, from reflection off the NS surface, the accretion disk, the disk wind, or the stellar wind of the companion star \citep[see][for discussion]{2022-cenx3}. 
This may lead to these polarized components being entangled with the polarized emission produced at the NS  magnetic poles, resulting in complications when interpreting the behavior of the pulse-phase dependence of the polarization properties. 
Particularly, both bright transient XRPs \mbox{LS~V~+44~17} and \mbox{Swift~J0243.6+6124} required the addition of a second, constant (unpulsed) polarized component to explain their PA pulse-phase dependence  within the framework of the RVM, thereby allowing the determination of their geometrical  properties \citep{2023A&A...677A..57D,Poutanen24}.

\vel, a bright and persistent XRP, is often considered the archetypal wind accretor and remains one of the most studied HMXBs \citep{1967-Chodil}.
It is a binary system consisting of a NS and the stellar companion GP Vel, a B0.5Ib supergiant \citep{2003-quaintrell}.
\vel has a spin period of 283~s \citep{1976-McClintock} and an orbital period of 8.964~d \citep{1972-Ulmer,1995-van-Kerkwijk}, showing eclipses lasting for about 2~d, with a lower limit of the orbital inclination of $i=73\degr$ \citep{1995-van-Kerkwijk}.
\vel is deeply embedded in the stellar wind of its companion star, and the X-ray luminosity is known to be variable on all time-scales, with an average value of around $L_\mathrm{X}\sim4\times10^{36}$\,erg\,s$^{-1}$ \citep{2004-Staubert,2008-Kreykenbohm}.
\vel is known to display the so-called soft excess, which has been attributed to the contribution of an additional low-energy component to the overall spectral continuum.
The source of this low-energy component has not been confirmed, however, it has been hypothesized to originate from the dense stellar wind surrounding the NS, or from thermal emission at the NS surface (see, e.g., \citealt{2004-Hickox}).

Considering the results of the energy-resolved polarimetric analysis of \vel in Paper I and the well-established presence of a soft excess in the source spectrum, the energy-dependent nature of the polarization may be attributed to two separate polarized components of different origins. 
This results in the complex behavior displayed by the PA pulse-phase dependence. 
Consequently, the pulse-phase dependence of the PA at low and high energies should exhibit different behavior. 
The RVM can then be applied to the PA pulse-phase dependence of the component associated with the emission from the accretion region at the NS magnetic poles (i.e., the high-energy component).
Here, we attempt to disentangle the contribution of the two components to the energy dependence of \vel polarization properties by a detailed spectro-polarimetric analysis using observations made by \ixpe.

\section{Data} 
\label{sec:data}

\ixpe is an observatory launched in December 2021 as a NASA/ASI mission, providing imaging polarimetry over the 2--8 keV energy range \citep{Weisskopf2022}.
\ixpe consists of three grazing incidence telescopes, each consisting of a mirror module assembly (MMA), which focuses X-rays onto a focal-plane polarization-sensitive gas pixel detector unit (DU).
In addition to measuring the sky coordinates, time of arrival, and energy of each detected photon, it also measures the direction of the photo-electron, which allows for performing polarimetric analyses.

The results of the analysis of the \ixpe data are published in Paper I, based on two observations of \vel conducted by \ixpe (01002501, 02005801) between 2022 April 15--21 and November 30--December 6 with the total effective exposures of 280 and 270~ks, respectively.
The following analysis is based on the same (level-2) data set, which has been downloaded from the HEASARC archive,\footnote{\url{https://heasarc.gsfc.nasa.gov/cgi-bin/W3Browse/w3browse.pl}} and processed with the {\sc ixpeobssim} package version 30.2.1 using the CalDB version 20211209:v13 for observation 1, and for observation 2, using the CalDB version 20220702:v13.

The source photons were extracted from a region of radius $R_{\rm src}=70\arcsec$.
We use the \texttt{barycorr} tool from the \textsc{ftools} package to correct event arrival times to the barycenter of the solar system.
To correct for orbital motion in a binary system, we use the orbital parameters for \vel provided by the Fermi Gamma-ray Burst Monitor.\footnote{\url{https://gammaray.nsstc.nasa.gov/gbm/science/pulsars.html}}
Because of the high count rate from the source, we do not subtract the background \citep{Di_Marco_2023}.
Considering the similarities between the two \ixpe observations, we combine the data into one single set, excluding the eclipse data from observation 1. 
We use the information provided in Paper I to phase-tag the events, with the phase difference between the pulse profiles from the first and second observations determined by cross-correlating the profiles (using the implementation provided by the Python library \textsc{NumPy}).

For the spectro-polarimetric analysis, source $I$, $Q$, and $U$ Stokes spectra were extracted using the \texttt{xpbin} tool's \texttt{PHA1}, \texttt{PHA1Q}, and \texttt{PHA1U} algorithms, producing a data set of nine spectra per observation, three for each DU.
The $I$, $Q$, and $U$ spectra from observations 1 and 2 were subsequently combined using the \texttt{mathpha} tool.
Corresponding response files for the observations were combined using the \texttt{addrmf} and \texttt{addarf} tools.
Finally, the subsequent spectro-polarimetric analysis is performed using \textsc{xspec} package \citep{Arn96}, version 12.14.0. 
Stokes $I$ spectra have been binned to have a minimum of 30 counts per energy channel. The same energy binning was subsequently applied to the Stokes $Q$ and $U$ spectra. The total set of nine spectra was fitted with \textsc{xspec} using $\chi^2$ statistics, and model parameter uncertainties are presented at the 68.3\% confidence level (1$\sigma$) unless stated otherwise.

\section{Results} 
\label{sec:res}

\subsection{Spectral model with two polarized components}

Some XRPs display a soft component in their spectra, in addition to the power-law component, which is frequently modeled as a blackbody or thermal bremsstrahlung \citep{2004-Hickox}.
These spectral components, i.e., hard and soft components, may have entirely separate origins.
The hard (power-law) component is associated with the accretion region at the NS surface, while the soft component may have a different origin (see Sect.~\ref{sec:spec-comps}), which may or may not display pulse phase dependence.
A spectro-polarimetric analysis extracting the individually polarized properties of the soft and hard components can be utilized to disentangle the properties of the spectral continuum. 

%%%%%%%%%%%%%%%%%%%%%%%%%%%%%%%%%%%%%%%%%%%%%%%%%%%
\begin{figure}
\centering
\includegraphics[width=0.85\linewidth]{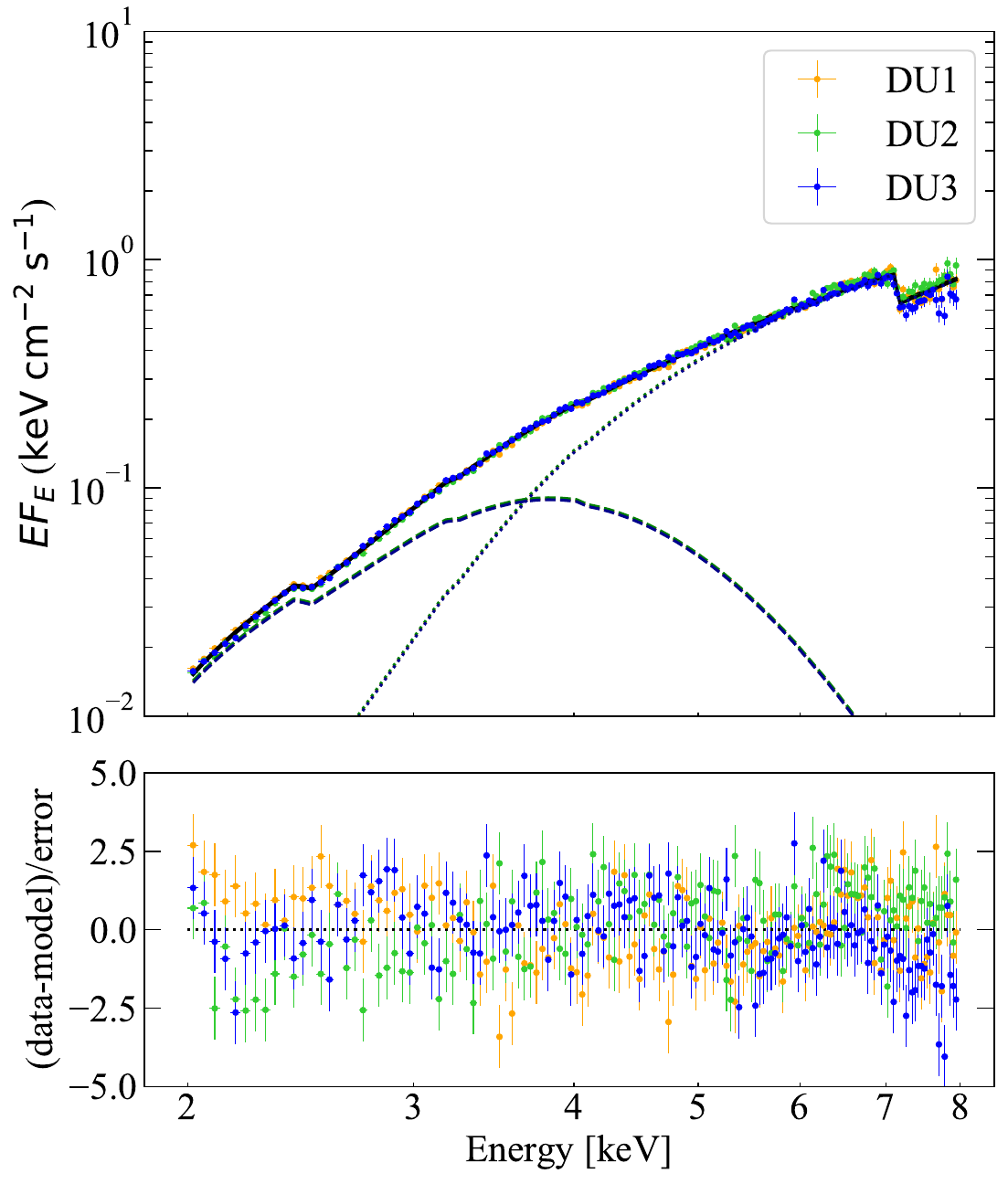}
\caption{Unfolded energy spectra (Stokes $I$ parameter) of \vel in $EF_E$ representation, using a two-component model fit to the \ixpe data set.  
The solid, dashed, and dotted lines show the total model spectrum, the thermal bremsstrahlung component, and the power-law component, respectively. 
The bottom panel shows the fit residuals. 
}
\label{fig:two-comp-spectra}
\end{figure}
%%%%%%%%%%%%%%%%%%%%%%%%%
%%%%%%%%%%%%%%%%%%%%%%%%%%%%%%%%%%%%%%%%%%%%%%%%%%%
\begin{figure}
\centering
\includegraphics[width=0.75\linewidth]{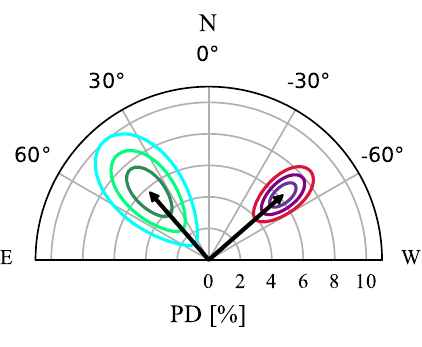}
\caption{Polarization vector of \vel from the results of the phase-averaged spectro-polarimetric analysis of the  \ixpe data set. Contours at 68.3\%, 95.45\% and 99.73\% confidence are shown for the \texttt{bremss} and \texttt{powerlaw} components in greenish and reddish colors, respectively. }
\label{fig:two-comp-phase-ave}
\end{figure}
%%%%%%%%%%%%%%%%%%%%%%%%%

%%%%%%%%%%%%%%%%%%%%%%%%%%%%%%%%%%%%%%%%%%%%%%%%%%%
\begin{figure*}
\centering
\includegraphics[width=0.4\textwidth]{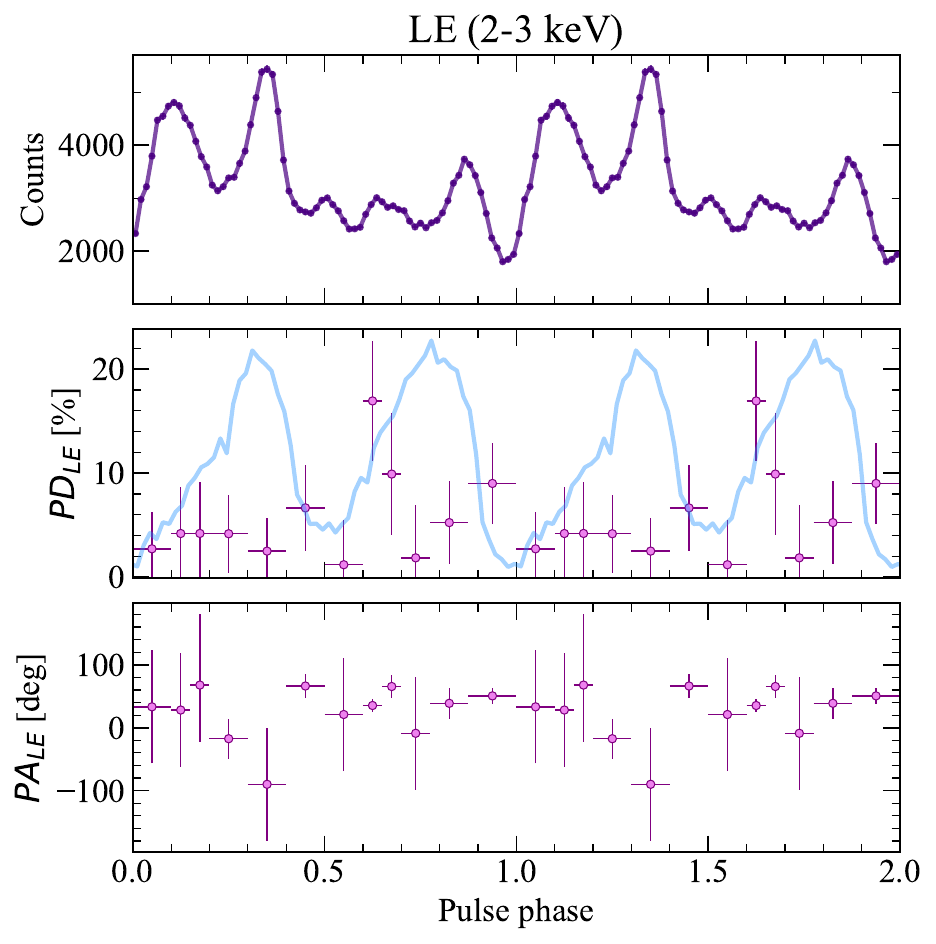}
\includegraphics[width=0.4\textwidth]{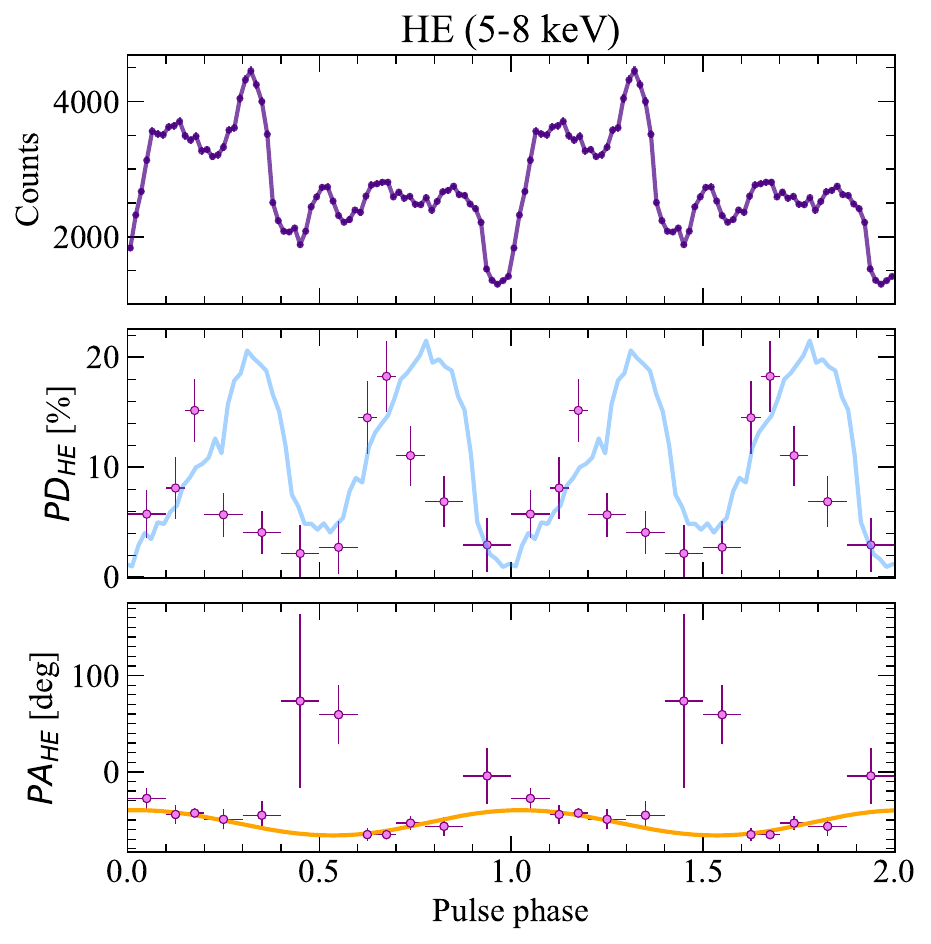}
\caption{Results of the phase-resolved spectro-polarimetric analysis of \vel for the \ixpe data in the LE (2--3 keV; \textit{left}) and HE (5--8 keV; \textit{right}).
\textit{Top panels:} The pulse profiles (in counts) corresponding to the LE and HE ranges. \textit{Middle} and \textit{bottom panels} demonstrate the pulse-phase dependence of the PD and PA, respectively.
The NuSTAR pulse profile in the 15--45 keV energy band is displayed in the PD panels (light blue).
The orange curve in the PA panel shows the best-fit RVM to the data in the HE range.}
\label{fig:two-comp-HELE}
\end{figure*}
%%%%%%%%%%%%%%%%%%%%%%%%%

%%%%%%%%%%%%%%%%%%%%%%%%%%
\begin{table}
\centering
\caption{Best-fit spectro-polarimetric parameters of model \eqref{eq:model} for the data set.
}
\begin{tabular}{llcc}
\hline
\hline
 Component & Parameter & Unit & Value   \\ 
\hline
        \texttt{tbabs} & $N_{\mathrm{H}}$ & $10^{22}\mathrm{\;cm^{-2}}$  & $10.0\pm0.5$ \\%& $9.8\pm0.02$ \\
        \texttt{tbpcf} & $N_{\mathrm{H,tbpcf}}$ & $10^{22}\mathrm{\;cm^{-2}}$ & $28.2_{-1.1}^{+1.3}$ \\%& $27.5\pm0.01$ \\
        & $f_{\mathrm{cov}}$  &   & $0.94\pm0.01$ \\%& $0.94\pm0.002$ \\
        \texttt{powerlaw} & $\Gamma$  &   & $1.08\pm0.06$ \\%& $1.04\pm0.02$ \\
        \texttt{bremss} & $kT$ & keV  & $0.70\pm0.04$ \\%& $0.7^{\mathrm{fixed}}$ \\
        & norm   &   &  $33_{-10}^{+12}$ \\%& $32.8^{\mathrm{fixed}}$ \\
        \texttt{constant} & $\mathrm{const_{DU2}}$  &  & $1.014\pm0.003$ \\%& $1.013\pm0.003$ \\
        & $\mathrm{const_{DU3}}$  &  & $0.994\pm0.003$ \\%& $0.994\pm0.003$ \\
        \texttt{polconst} & $\mathrm{PD_{bremss}}$ & \%  & $5.4\pm1.4$ \\%& -- \\
        & $\mathrm{PA_{bremss}}$ & deg & $41\pm7$ \\%& -- \\  
        \texttt{polconst} & $\mathrm{PD_{po}}$ & \%  & $6.2\pm0.7$ \\%& -- \\
        & $\mathrm{PA_{po}}$ & deg & $-49\pm3$ \\%& -- \\         
        \hline
        & $\chi^2$ (d.o.f.) & & 1566 (1328) \\%& 1561 (1328) \\ 
        \hline
\end{tabular}
\tablefoot{Uncertainties  computed using the \texttt{error} command are given at the 68.3\% (1$\sigma$) confidence level ($\Delta\chi^2=1$ for one parameter of interest). } 
\label{table:best-fit-twocomp}
\end{table}
%%%%%%%%%%%%%%%%%%%%%%%%%%

Several phenomenological models have been utilized to describe \vel spectral continuum (see \citealt{2021-Kretschmar}, and references therein).
The X-ray emission below 10 keV can be well described by a simple absorbed power law with an iron line at 6.4 keV.
In addition to this, \vel is well known to display a soft excess below 3 keV \citep{1994-Haberl}.
Due to the restricted energy range covered by IXPE and the energy resolution of the instrument \citep{Weisskopf2022}, we use a simplified model and do not include an iron line. Including an iron line at 6.4 keV results in a normalization of the line component that is consistent with zero (the other parameters are consistent whether the iron line is included or not). Hence, we do not include this component in the final model (neither in the phase-averaged nor the phase-resolved analysis).

The high-energy (HE) component was fitted with a simple power-law model \texttt{powerlaw}.
In order to account for the soft excess below 3~keV, we fit this low-energy (LE) component with the \texttt{bremss} model in \textsc{xspec} ($kT\sim0.5$~keV; \citealt{1994-Haberl}).
The interstellar absorption affecting the continuum was introduced using the model \texttt{tbabs} with the abundances from \citet{Wilms2000}. 
This was combined with the \texttt{polconst} polarization model, which assumes energy-independent PD and PA, applied to the individual components (\texttt{bremss} and  \texttt{powerlaw} components).
%\mg{The PA is defined in the $0\degr$ to $-180\degr$ interval.}
In order to achieve an acceptable fit, a partial covering absorption model \texttt{tbpcf} was introduced as well, which applies an added column density to a fraction of the continuum model \citep{2021-Kretschmar}.
The re-normalization constant, \texttt{const}, was used to account for the possibility of discrepancies between the different DUs, and for DU1 it was fixed to unity. 
The final spectral model 
\begin{eqnarray}\label{eq:model} 
 && \texttt{tbabs$\times$tbpcf$\times$(polconst$\times$bremss} \nonumber \\
&+&\texttt{polconst$\times$powerlaw$)\times$const} 
\label{eq:fit}
\end{eqnarray}
was used as the model for the initial phase-averaged spectro-polarimetric fitting.

The results of the phase-averaged spectro-polarimetric analysis using two separately polarized emission components are given in Table~\ref{table:best-fit-twocomp} and the $I$ (flux) spectra are shown in Fig.~\ref{fig:two-comp-spectra}.
The \texttt{steppar} command in \textsc{xspec} was used to produce confidence contours and the contour plots for the PD and PA are shown in Fig.~\ref{fig:two-comp-phase-ave}.
The results demonstrate that the two-component spectral model effectively explains the findings of the phase-averaged energy-resolved analysis.
Namely, the difference in PA between the HE and LE components is close to 90\degr. 
Figure~\ref{fig:two-comp-spectra} shows that the contribution of the LE and HE components at $\sim$3.5~keV is equal. 
Around this energy, the PD is zero, as the contribution from the two components with nearly equal PD cancels out. 
This is also in line with what was observed in the phase-averaged energy-resolved analysis (see Paper I). 

Ideally, we would like to examine the phase-resolved energy dependence of the polarization properties.
However, we do not have enough statistics to perform a detailed phase-resolved spectro-polarimetric analysis using a model with two separately polarized components.
Therefore, below we adopt a simplified approach by dividing the total energy range into LE and HE parts and use simple models for the spectral fitting. 

%%%%%%%%%%%%%%%%%%%%%%%%%
\begin{table*} 
\centering
\caption{Spectro-polarimetric parameters in different pulse-phase bins for the low- and high-energy components. }
\footnotesize 
\begin{tabular}{ccccccccccc} 
    \hline\hline
    Phase & $q_{\mathrm{LE}}$ & $u_{\mathrm{LE}}$ &  PD$_{\mathrm{LE}}$ & PA$_{\mathrm{LE}}$ & $\chi^2$/d.o.f. & $q_{\mathrm{HE}}$ & $u_{\mathrm{HE}}$ & PD$_{\mathrm{HE}}$ & PA$_{\mathrm{HE}}$ & $\chi^2$/d.o.f. \\
          & (\%) & (\%) & (\%) & (deg) & & (\%) & (\%) & (\%) & (deg) &  \\ 
    \hline
    0.000--0.100 & $2.3\pm3.7$ & $3.5\pm3.7$ & $2.7_{-2.7}^{+3.6}$ & $33\pm90$ & 245/211 & $2.6\pm2.6$ & $-5.1\pm2.6$ & $5.7\pm2.2$ & $-28\pm11$ & 653/584 \\
    0.100--0.150 & $-0.1\pm4.6$ & $2.2\pm4.6$ & $4.2_{-4.2}^{+4.5}$ & $28\pm90$ & 239/211 & $-2.0\pm3.3$ & $-10.1\pm3.3$ & $8.1\pm2.8$ & $-44\pm10$ & 547/518 \\   
    0.150--0.200 & $-2.7\pm5.1$ & $4.0\pm5.1$ & $4.2_{-4.2}^{+4.9}$ & $68\pm90$ & 225/211 & $1.1\pm3.5$ & $-20.0\pm3.5$ & $15.2\pm2.9$ & $-43\pm5$ & 534/518 \\  
    0.200--0.300 & $3.8\pm3.8$ & $0.4\pm3.8$ & $4.2\pm3.7$ & $-18\pm32$ & 221/211 & $0.9\pm2.4$ & $-6.0\pm2.4$ & $5.7\pm2.0$ & $-49\pm10$ & 613/617 \\  
    0.300--0.400 & $-2.1\pm3.2$ & $-1.0\pm3.2$ & $2.5_{-2.5}^{+3.1}$ & $-89\pm90$ & 231/211 & $0.1\pm2.4$ & $-4.5\pm2.4$ & $4.1\pm2.0$ & $-45\pm14$ & 786/605 \\   
    0.400--0.500 & $-1.4\pm4.2$ & $5.1\pm4.2$ & $6.7\pm4.1$ & $66\pm19$ & 234/211 & $-3.0\pm3.0$ & $1.6\pm3.0$ & $2.3_{-2.3}^{+2.5}$ & $73\pm90$ & 612/560 \\  
    0.500--0.600 & $2.4\pm4.4$ & $4.3\pm4.4$ & $1.2_{-1.2}^{+4.3}$ & $21\pm90$ & 256/211 & $1.0\pm2.9$ & $3.7\pm2.9$ & $2.7\pm2.4$ & $59\pm31$ & 592/566 \\
    0.600--0.650 & $4.5\pm5.9$ & $18.0\pm5.9$ & $17.0\pm5.8$ & $35\pm10$ & 209/211 & $-11.3\pm4.0$ & $-12.7\pm4.0$ & $14.5\pm3.3$ & $-65\pm7$ & 510/488 \\ 
    0.650--0.700 & $-4.9\pm6.0$ & $6.1\pm6.0$ & $9.9\pm5.8$ & $65\pm18$ & 204/211 & $-13.0\pm3.8$ & $-18.6\pm3.8$ & $18.3\pm3.2$ & $-65\pm5$ & 526/497 \\ 
    0.700--0.775 & $1.0\pm5.2$ & $-0.1\pm5.2$ & $1.9_{-1.9}^{+5.0}$ & $-9\pm90$ & 202/211 & $-3.6\pm3.2$ & $-11.4\pm3.2$ & $11.1\pm2.7$ & $-53\pm7$ & 558/548 \\ 
    0.775--0.875 & $2.4\pm4.1$ & $5.8\pm4.1$ & $5.3\pm4.0$ & $39\pm25$ & 188/211 & $-2.0\pm2.8$ & $-5.8\pm2.8$ & $6.9\pm2.3$ & $-57\pm10$ & 552/578 \\ 
    0.875--1.000 & $-0.3\pm4.0$ & $8.5\pm4.0$ & $9.0\pm3.9$ & $51\pm13$ & 258/211 & $2.6\pm3.0$ & $-0.4\pm3.0$ & $2.9\pm2.5$ & $-4\pm29$ & 607/557 \\ 
    \hline
    \end{tabular}
\tablefoot{
PD and PA are obtained with \textsc{xspec}. 
Normalized Stokes parameters $q$ and $u$ are obtained using the \texttt{pcube} algorithm in {\sc ixpeobssim}.   
The uncertainties computed using the \texttt{error} command are given at the 68.3\% (1$\sigma$) confidence level ($\Delta\chi^2=1$ for one parameter of interest). 
}
\label{table:best-fit-twocomp-phaseres}
\end{table*}
%%%%%%%%%%%%%%%%%%%%%%%%%%

\subsection{Analysis of low- and high-energy emission}

Utilizing the results from the phase-averaged analysis with two polarized components, we confine the fit to LE and HE separately.
From the point where the contributions from the LE and HE components to the total polarized flux are equal (i.e. where the PD is equal to zero), we determine the break-off point between the LE- and HE-ranges.
To minimize the contribution of different spectral components to the nearby band, the 3--5 keV range was completely excluded from the analysis.
The final energy ranges used for the phase-resolved spectro-polarimetric fitting (with minimal contribution from the other component) for the LE and HE are 2--3  and 5--8~keV, respectively.
The values of the \texttt{tbabs} and \texttt{tbpcf} parameters were fixed to that of the phase-averaged results.

For the spectro-polarimetric analysis at LE, we used a simplified model:
\begin{equation} \label{eq:fit-LE}
\texttt{tbabs$\times$tbpcf$\times$(polconst$\times$bremss)$\times$const} ,
\end{equation}
which was applied to each phase bin in the phase-resolved data set.
The results of the phase-resolved spectro-polarimetric fit in the LE energy range are given in Table~\ref{table:best-fit-twocomp-phaseres} and shown in the left panels of Fig.~\ref{fig:two-comp-HELE}.
Normalized Stokes $q$ and $u$ parameters are extracted using the \texttt{xpbin} tool's \texttt{pcube} algorithm included in the {\sc ixpeobssim} package, which has been implemented according to the formalism by \citet{2015APh....68...45K}. 
However, low counting statistics make it impossible to constrain the behavior of the PD and PA.
For the \texttt{polconst} model, the PD is consistent with zero for most of the phase bins, and the PA is unconstrained.
However, for the few number of phase bins where we can significantly detect the PD, we can compare the results with the ones from the HE analysis (see below).

%%%%%%%%
\begin{figure}
\centering
\includegraphics[width=0.75\linewidth]{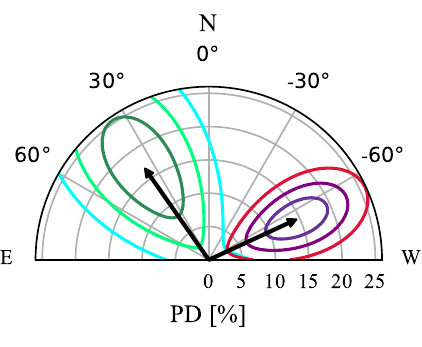}
\caption{Polarization vectors of \vel from the results of the phase-resolved spectro-polarimetric analysis for the phase interval 0.600--0.650. Contours at 68.3\%, 95.45\%, and 99.73\% confidence are shown for the \texttt{bremss} and \texttt{powerlaw} components in greenish and reddish colors, respectively.}
\label{fig:combcontours-bin8}
\end{figure}
%%%%%%%%%%%%%%%%%%%%%%%%%

The phase-resolved spectro-polarimetric analysis of the HE component (5--8 keV) was performed using the model:
\begin{equation} \label{eq:fit-HE}
\texttt{tbabs$\times$tbpcf$\times$(polconst$\times$powerlaw)$\times$const}.
\end{equation}
The results are given in Table \ref{table:best-fit-twocomp-phaseres} and shown in the right panels of Fig.~\ref{fig:two-comp-HELE}.
As can be seen from the figure, this results in a nicely structured behavior of both the PD and PA, where the pulse-phase dependence of the PA displays a single sine wave over the pulse phase (compared to the complicated behavior of the PA in Paper I).
Therefore, we can use these results to model the PA phase dependence with the RVM model and determine the pulsar geometry for \vel for the first time.
It is worth noting that different high-energy ranges (3.5--8, 4--8, and 4.5--8 keV) all give us consistent results for the HE phase-dependence of the PA.

Due to the fact that we observe a 90\degr\ difference in the PA between the LE and HE component in the phase-averaged analysis, we would like to examine if the same behavior is observable in the phase-resolved analysis. 
This can be achieved by comparing the contour plots of the LE and HE components in each phase bin.
We observe significant polarization in phase bin 0.600--0.650, with the only highly significant PD detection for the LE also in this phase interval.
Comparing the PA of the LE and HE components in this phase bin, we see a difference in the PA of close to 90\degr\ (see Fig.~\ref{fig:combcontours-bin8} for combined contour plots of the LE and the HE). 
For the remaining phase bins, the differences between the PA of the HE and the LE components are consistent with 90\degr, with the PA unconstrained for a large number of phase intervals for the LE.

The pulse profile of \vel is well known to show both a unique and complex five-peaked structure at lower energies.
However, the pulse profile evolves to a simpler two-peak structure at higher energies.
Using NuSTAR data in the 15--45 keV energy range, we compare the high-energy pulse profile to the results of the phase-resolved spectro-polarimetric analysis, for both the LE and HE polarization results (see Fig.~\ref{fig:two-comp-HELE}).
The pulse phase dependence of the PD shows a noticeable correlation (albeit with some shift in the phase) with the high-energy pulse profile, with the flux minima roughly corresponding to the minima in the PD.

\section{Discussion} 
\label{sec:discussion}

The observations of XRPs made by \ixpe have revealed a more complicated picture of these sources in terms of their polarization properties than what has been predicted by theory. 
Significantly lower PDs than expected provide an incentive to revisit theoretical models and consider different mechanisms that may contribute. 
Several possible processes that may be involved in producing the polarization observed in XRPs were discussed in \cite{2022-cenx3}.
Consequently, the observed polarization properties may in fact be affected by a multitude of different processes in the immediate environment of the NS, as well as processes related to the interaction of photons under the extreme conditions of a highly magnetized plasma. 
These processes may in turn lead to a more complex spectrum, where contributions from different components arise and require their own interpretation.

Considering the complex nature of the pulse-phase dependent polarization properties of \vel, it is likely that the interplay of several factors may be responsible. 
Below, we discuss possible explanations for \vel unique energy dependent polarimetric behavior in terms of two separate mechanisms. First, we consider the possibility that the polarization may be attributed to two independently polarized spectral components, with a 90\degr\ difference in PA between them.
For the second mechanism, the 90\degr\ difference in the PA is considered to be introduced by vacuum resonance, where a conversion from one polarization mode to the other occurs.
Finally, the disentanglement of the pulse-phase dependent polarization properties of the HE component from that of the LE component offers us the opportunity to determine the geometrical parameters of the pulsar.

\subsection{Two spectral components}
\label{sec:spec-comps}

The existence of soft excess in \vel is well established and, as mentioned in the introduction, different origins to it have been proposed but have yet to be confirmed.
The soft excess appears as an additional contribution at low energies to the hard spectral continuum.
The general spectral continuum of \vel can be described using two components: a power law and a thermal bremsstrahlung, which accounts for the additional low energy flux from the source.
The hard (power-law) spectral component of emission from \vel is believed to originate from the accretion region, i.e. the accretion mound, at the NS magnetic poles, while the origin of the soft component is not known, although several possibilities have been hypothesized.
\citet{2004-Hickox} provided an extensive review of the possible origin of soft excesses in XRPs in general and in \vel in particular.
For \vel, the soft excess is believed to either originate from the stellar wind surrounding the NS, or thermal emission from the NS surface.

It is difficult to draw any conclusions about the origin of the soft component based on its polarization properties.
Because of the fact that we observe a similar level of pulsed fraction in the LE band as in the HE band, we can conclude that the LE component is pulsed and should originate from a sufficiently compact region.
Therefore, it is possible that the LE component is produced by thermal emission from compact regions at the NS surface, in the immediate vicinity of the base of the accretion mound and illuminated by it, resulting in a 90\degr\ difference in PA compared to that of the HE component.

%%%%%%%%%%%%%%%%%%%%%%%%%%%%%%%%%%%%%%%%%%%%%%
\subsection{Vacuum polarization}
%%%%%%%%%%%%%%%%%%%%%%%%%%%%%%%%%%%%%%%%%%%%%%

\vel is a relatively low-luminosity XRP operating in the sub-critical accretion regime \citep[see, e.g.,][]{1976MNRAS.175..395B,2015MNRAS.447.1847M}.  
The polarization of its radiation can thus be considered within the framework of a toy model for an accretion-heated NS atmosphere, initially proposed for Her X-1 \citep{2022NatAs...6.1433D}.  
In this model, the upper atmospheric layers are overheated by plasma deceleration via Coulomb collisions \citep[e.g.,][]{1969SvA....13..175Z,2000ApJ...537..387Z, 2018A&A...619A.114S,2019MNRAS.483..599G}.  

Let us consider a toy model describing the upper layers of the neutron star atmosphere heated by the accreting particles. We assume that the base atmosphere is isothermal and the upper layers are overheated.  
We take a predefined temperature dependence on the column density $m$ in the atmosphere:  
\begin{equation}  
T(m) = (T_{\rm up} - T_{\rm low})\exp(-m/m_{\rm up}) + T_{\rm low},  
\end{equation}  
where $T_{\rm up}$ is the temperature of the upper overheated layer, $T_{\rm low}$ is the temperature of the underlying atmosphere, and $m_{\rm up}$ is the transition column density between these two layers.  
We consider a fully ionized pure hydrogen atmosphere with the gas pressure $P = mg$ determined from the integral of the hydrostatic equilibrium equation, where  $g$ is the surface gravity.  
For simplicity, we ignore the external ram pressure of the accretion flow as well as the radiation pressure.   
Next, we solve the radiation transfer equation in the two polarization modes with the standard boundary conditions.
The corresponding opacities, including vacuum polarization and mode conversion at the vacuum resonance, are incorporated following \citet{2006MNRAS.373.1495V}. 
The code is described in detail in \citet{2009A&A...500..891S}.  

For the model calculations, we assumed a NS mass of 1.4~$M_\odot$ and a radius of $R = 12$\,km, with a surface magnetic field strength of $B = 3 \times 10^{12}$\,G, which corresponds to the energy of the observed cyclotron feature in \vel.  
We found that the polarization of the emergent radiation in the energy range 2--8~keV can be sufficiently low at certain model parameters, as was previously observed for \mbox{Her X-1} \citep{2022NatAs...6.1433D}. 
Here, we consider a single model with the following parameters: $T_{\rm up} = 6 \times 10^8$\,K, $T_{\rm low} = 10^7$\,K, and $m_{\rm up} = 0.75$\,g\,cm$^{-2}$.  

%%%%%%%%%%%%%%%%%%%%%%%%%%
\begin{figure} 
\centering
\includegraphics[width=0.9\linewidth]{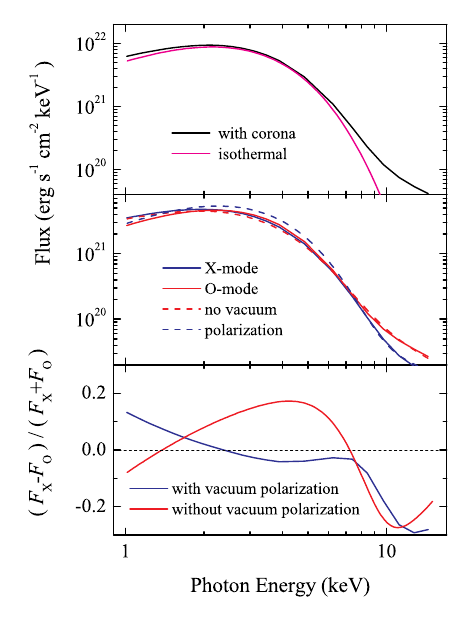}
\caption{Energy dependencies of the fluxes and polarization. 
Upper panel: Emergent spectra from the isothermal atmosphere (red curve) and from the atmosphere with the hot upper layer (black curve). 
Middle panel: Emergent spectra in the normal modes for the model with the hot upper layer with (solid curves) and without (dashed curves) vacuum polarization taken into account.
Bottom panel: Energy dependence of the PD for the hot layer model with (blue curve) and without (red curve) vacuum polarization taken into account. }
\label{fig:sppol}
\end{figure}
%%%%%%%%%%%%%%%%%%%%%%%

\begin{figure} 
\centering
\includegraphics[width=0.9\linewidth]{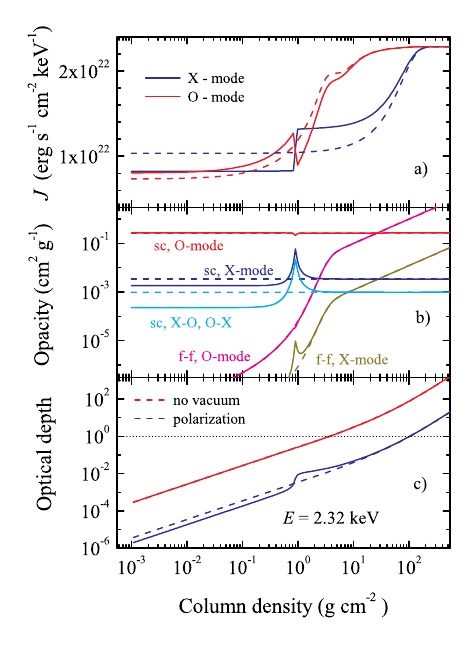}
\caption{Dependencies of the mean intensities (upper panel), scattering and free-free opacities (middle panel), and the optical thickness (bottom panel) of the normal modes on the column density $m$ with (solid curves) and without (dashed curves) vacuum polarization taken into account.}
\label{fig:int}
\end{figure}

\begin{figure} 
\centering
\includegraphics[width=0.9\linewidth]{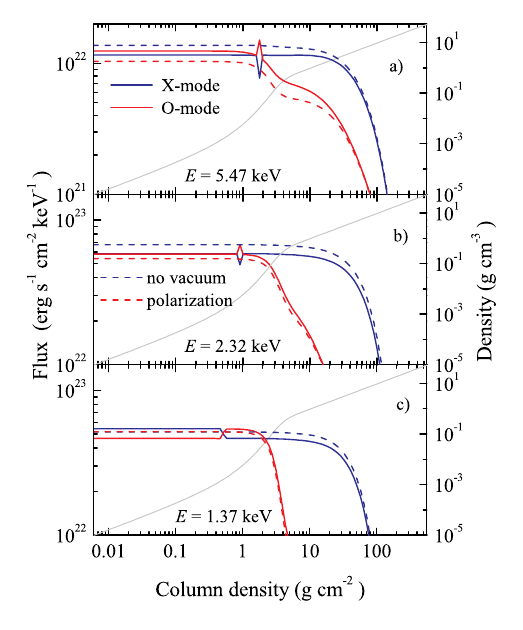}
\caption{Dependencies of the fluxes  in the normal modes at different photon energies on column density $m$ with (solid curves) and without (dashed curves) vacuum polarization taken into account.
The mass density is also shown with the light gray solid curves (right axes).}
\label{fig:modecheng}
\end{figure}

The results of the radiation transfer computations in two  normal modes for this model are presented in Figs.\,
\ref{fig:sppol},  \ref{fig:int}, and \ref{fig:modecheng}. The emergent spectrum of the model is compared with the model spectrum of the isothermal atmosphere (upper panel of Fig.\,\ref{fig:sppol}). It is clear that the upper overheated layer adds the optically thin emission to the thermal blackbody-like radiation of the isothermal atmosphere. This optically thin radiation is mainly emitted in the O-mode because the absorption (free-free) opacity is larger in this mode (see the middle panels in Figs.\,\ref{fig:sppol} and \ref{fig:int}). It is interesting that the vacuum polarization is insignificant at the photon energies above 8 keV, and converts X- and O-modes to each other at low energies below 2 keV. And there is a relatively broad energy region 2--8 keV where vacuum polarization makes the intensities in X- and O-modes close to each other resulting in a close to zero PD (Fig.\,\ref{fig:sppol}, bottom panel). 
These results are connected to the vacuum polarization properties, which we discuss below.

%%%%%%%%%%%%%%%%%%%%%%%%%%
%%%%%%%%%%%%%%%%%%%%%%%%%

The importance of vacuum polarization for radiation transfer in highly magnetized model atmospheres of neutron stars was investigated in detail by \citet{2002ApJ...566..373L, 2003ApJ...588..962L, 2003PhRvL..91g1101L}. 
The main conclusions are the following. 
Virtual electron-positron pairs in a strong magnetic field change the dielectric tensor of the plasma and, as a result, the plasma opacity, especially in the X-mode (see Fig.\,\ref{fig:int}, middle panel). 
The opacity increases significantly at the so-called vacuum resonance depth and in the deeper layers, whereas in the more upper layers, it decreases. 
As a result, the optical thickness over the whole atmosphere changes (Fig.\,\ref{fig:int}, bottom panel) which leads to different distributions of mean intensities over depth (Fig.\,\ref{fig:int}, top panel). 
The intensities in the two modes are partially converted to each other at the vacuum resonance depth. 
We show in Fig.\,\ref{fig:int} the results obtained for the photon energy at which the vacuum resonance produces an equality of intensities of the two modes. 

The vacuum resonance occurs at the depth where the contributions of the vacuum polarization and the plasma to the dielectric tensor become equal. 
The plasma density at this depth must satisfy the following condition
\begin{equation} 
\rho_{\rm V} \approx E^2\,B_{14}^2 \, \text{g\,cm}^{-3},
\end{equation} 
where $E$ is the photon energy in keV  and $B_{14} = B/10^{14}$\,G.   
It means that for the given magnetic field strength the depth of the vacuum resonance depends strongly on the photon energy.

The modes are almost completely converted to each other after passing the vacuum resonance. At the low photon energies, the vacuum resonance takes place in the upper layers which are optically thin in both modes.
In this case the emergent fluxes in the modes just also converted to each other  (see Fig.\,\ref{fig:modecheng}, bottom panel). If the vacuum resonance occurs in the deep layers where both modes are thermalized with equal intensities, its presence does not affect the spectrum of the emergent radiation in any way.

The most interesting case is when the vacuum resonance takes place at the depth where the X-mode has already formed and is in an optically thin state, whereas the radiation in O-mode is still in the process of formation. 
Then the intensity in the X-mode drops to the O-mode level and stays constant. 
Whereas the intensity in O-mode increases to the level of X-mode intensity, but then continues to decrease with depth (see Fig.\,\ref{fig:int}, top panel). 
Therefore, at some photon energy, these intensities and the emergent fluxes can become equal to each other (see Fig.\,\ref{fig:modecheng}, middle panel). 
In a standard atmosphere, the density decreases smoothly toward the surface, and the described situation occurs in a narrow range of photon energies.
But in the transition layer between the normal atmosphere and the overheated layer, the density gradient is significant and the emergent fluxes in both modes could be close to each other in the relatively broad photon energies (see top panel of Fig.\,\ref{fig:modecheng} and Fig.\,\ref{fig:int}).

%%%%%%%%%%%%%%%%%%%%%%%%%%
%%%%%%%%%%%%%%%%%%%%%%%%%

We do not pretend to explain the whole X-ray energy spectrum of Vela X-1 using this toy model. The cyclotron emission together with the Compton scattering have to be included for that. We just demonstrate that the overheated upper layers of the atmosphere and the transition layer can create conditions under which the polarization of the emergent radiation will be close to zero in a relatively wide range of energies below 10 keV. We also note that the suggested toy model is a variation of the two-component model discussed in the previous subsection. The change of the polarization at the energies below 2.5 keV occurs due to the contribution of the optically thin emission of the overheated layer.

%%%%%%%%%%%%%%%%%%%%%%%%%%%%%%%%%%%%%%%%%%%%%%%%%%%
\begin{figure*}
\centering
\includegraphics[width=0.7\linewidth]{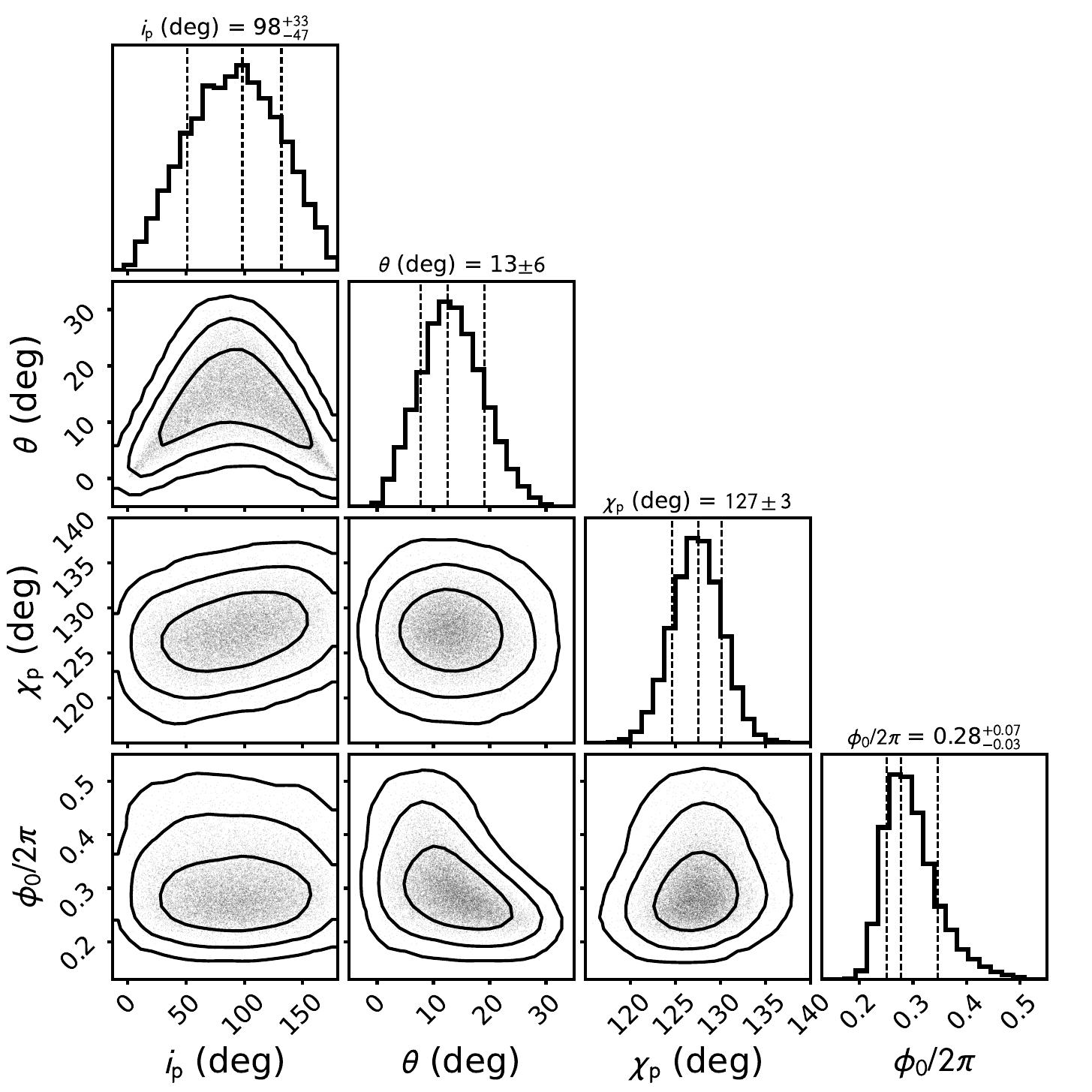}
\caption{Corner plot of the posterior distribution for the parameters of the RVM model fitted to the PA values obtained from the phase-resolved spectro-polarimetric analysis of the HE component. 
The two-dimensional contours correspond to 68.3\%, 95.45\% and 99.73\%  confidence levels. 
The histograms show the normalized one-dimensional distributions for a given parameter derived from the posterior samples.
The vertical dashed lines represent the mode of the distribution (central line) and the lower and upper quartiles (68\% confidence interval). The mode is calculated using a Gaussian kernel density estimate with a bandwidth of 0.1.}
\label{fig:rvm}
\end{figure*}
%%%%%%%%%%%%%%%%%%%%%%%%%

%%%%%%%%%%%%%%%%%%%%%%%%%%%%%%%%%%%%%%%%%%%%%%%%%%
\subsection{Determination of the pulsar geometry}
%%%%%%%%%%%%%%%%%%%%%%%%%%%%%%%%%%%%%%%%%%%%%%%%%%

The RVM can be used to constrain the pulsar geometry.
The geometrical properties of several XRPs observed by \ixpe have already been obtained \citep{2022NatAs...6.1433D,2022-cenx3,2023MNRAS.524.2004M,2023A&A...675A..48T,2023A&A...675A..29M,2023A&A...677A..57D,Suleimanov2023,Heyl24,2024MNRAS.tmp.1187Z,Poutanen24,Forsblom24}. 
If the radiation is assumed to be dominated by O-mode photons, the PA is (see Eq.\,(30) in \citealt{Poutanen2020}):
\begin{equation} \label{eq:pa_rvm}
\tan (\mbox{PA}\!-\!\chi_{\rm p})\!=\! \frac{-\sin \theta\ \sin (\phi-\phi_0)}
{\sin i_{\rm p} \cos \theta\!  - \! \cos i_{\rm p} \sin \theta  \cos (\phi\!-\!\phi_0) } ,
\end{equation} 
where $\chi_{\rm p}$ is the position angle (measured from north to east) of the pulsar angular momentum, $i_{\rm p}$ is the inclination of the pulsar spin to the line of sight, $\theta$ is the magnetic obliquity (i.e. the angle between the magnetic dipole and the spin axis), and $\phi_0$ is the phase when the northern magnetic pole passes in front of the observer. 
If instead, the radiation escapes predominantly in the X-mode, the position angle of the pulsar angular momentum is $\chi_{\rm p}\pm90\degr$.
The polarization plane actually rotates as the radiation travels through the NS magnetosphere, up to the adiabatic radius.
At such a distance, the dipole magnetic field component will dominate, and under these conditions, the RVM is applicable.

We can fit the RVM to the pulse-phase dependent PA obtained from the spectro-polarimetric analysis of the HE component using the affine invariant Markov Chain Monte Carlo (MCMC) ensemble sampler {\sc emcee} package of {\sc python} \citep{2013PASP..125..306F}. 
Because the PA is not normally distributed, we use the probability density function of the PA, $\psi$, from \citet{Naghizadeh1993}:
\begin{equation} \label{eq:PA_dist}
G(\psi) = \frac{1}{\sqrt{\pi}} 
\left\{  \frac{1}{\sqrt{\pi}}  + 
\eta {\rm e}^{\eta^2} 
\left[ 1 + {\rm erf}(\eta) \right]
\right\} {\rm e}^{-p_0^2/2}.
\end{equation}
Here, $p_0= p/\sigma_{\rm p}$ is the measured PD in units of the error,  $\eta=p_0 \cos[2(\psi-\psi_0)]/\sqrt{2}$, $\psi_0$ is the central PA obtained from \textsc{xspec}, and \mbox{erf} is the error function.
We applied the likelihood function $L= \Pi_i  G(\psi_i)$ with the product taken over all phase bins.  
The covariance plot for the parameters is shown in Fig.~\ref{fig:rvm}.   
The RVM provides an overall good fit to the PA of the HE component.
The magnetic obliquity and the pulsar position angle are accurately determined as $\theta=13\degr\pm6\degr$ and $\chi_{\rm p}=127\degr\pm3\degr$, respectively. The inclination is less well constrained, however, the estimate of $i_{\rm p}=98_{-47}^{+33}$~deg is consistent with the lower limit on the orbital inclination of 73\degr\ given by \citet{1995-van-Kerkwijk}.

\citet{1995-Bulik} found evidence for non-antipodal polar caps in \vel using a spectral model of an inhomogeneous magnetized atmosphere fitted to phase-resolved data obtained with \textit{Ginga}.
The rotation axis and the mean magnetic axis were found to be relatively close (i.e., they found small magnetic obliquities for both caps) and are seen at a large inclination with respect to the line of sight. The angle between the center of the polar cap and the rotational axis for caps 1 and 2 were found to be $22\degr\pm12\degr$ and $168\degr\pm12\degr$, respectively, for the most significant fits.
These results are in line with what is displayed in Fig.~\ref{fig:rvm}, where the results of the RVM-fitting similarly give us a small magnetic obliquity, with $\theta=13\degr\pm6\degr$.

\section{Summary}
\label{sec:sum}

\vel represents a unique case among the XRPs observed by \ixpe so far. 
\ixpe uncovered a clear energy dependence of its PD and PA, with the PD reaching zero at around 3.4 keV, corresponding to a switch in the PA of 90\degr. 
We have proposed an explanation for the curious case of \vel energy-dependent polarization properties. 
The results of our study can be summarized as follows:
\begin{enumerate}
\item The energy-dependence of the polarimetric properties of \vel can be explained by a two-component model, where the low-energy and high-energy components have different  polarization properties with the PA differing by 90\degr.
\item The phase-averaged polarimetric analysis using a spectral model consisting of two perpendicularly polarized components, a power-law component and a thermal bremsstrahlung component, effectively explains the energy-dependence of polarization including a 90\degr\ switch in PA between the components. 
At around 3.4 keV, the contribution from both components to polarized flux is equal, resulting in zero total PD, consistent with what was observed in Paper I.
\item The phase-resolved polarimetric analysis separating the low (2--3 keV) and high (5--8 keV) energies allows us to examine the phase-resolved polarization properties in these two different energy ranges. 
At low energies, the polarization in the phase-resolved data is predominantly unconstrained. 
At high energies, the PD and PA both display a nice structure over the pulse phase. 
The PD shows a correlation with the high-energy pulse profile, and the PA displays a single sine wave over the pulse period.
\item We propose two different scenarios to explain the 90\degr\  difference in PA between low and high energies.
The two-component spectral model assumes that the high-energy (power-law) component originates from the accretion mound, and the low-energy component originates from a compact region in the immediate vicinity of the base of the accretion mound.
Alternatively, the difference may arise due to the vacuum resonance in the NS atmosphere with an overheated upper layer. 
The effect of vacuum polarization on plasma opacity is most significant for those photon energies for which vacuum resonance occurs in the transition zone between the overheated upper and cold lower layers. As a result, at those relatively low photon energies ($E < 2.5$\,keV for our toy model) for which vacuum resonance occurs in the superheated upper layers of the atmosphere, the PD is positive (i.e. PA is parallel to the surface normal). 
At higher photon energies, the PD changes sign (i.e. PA is now perpendicular to the normal) because vacuum resonance occurs in the transition zone between the overheated and cold layers of the atmosphere.
Because the PA of the X-mode is 90\degr\ different from the PA of the O-mode, this gives an apparent switch in PA by 90\degr\ when going from low to high photon energies. 
\item We apply the RVM to the results of phase-resolved analysis at high energies to determine the pulsar geometry in \vel for the first time. 
The pulsar spin position angle and the magnetic obliquity are estimated as 127\degr\ and 13\degr, respectively, and the estimate for the inclination is consistent with previous estimates for the orbital inclination.
\end{enumerate}

%========================================
%========================================
\begin{acknowledgements}
The Imaging X-ray Polarimetry Explorer (IXPE) is a joint US and Italian mission. 
This research used data products provided by the IXPE Team and distributed with additional software tools by the High-Energy Astrophysics Science Archive Research Center (HEASARC), at NASA Goddard Space Flight Center.
This research has been supported by the Vilho, Yrjö, and Kalle Väisälä foundation (SVF), the Ministry of Science and Higher Education grant 075-15-2024-647 (SST, JP), the UKRI Stephen Hawking fellowship (AAM), and  Deutsche  Forschungsgemeinschaft (DFG) grant WE 1312/59-1 (VFS). 
\end{acknowledgements}

\bibliographystyle{aa}
\bibliography{allbib}

\end{document}